\documentclass[twocolumn,showpacs,preprintnumbers,amsmath,amssymb,nofootinbib]{revtex4}

\usepackage{graphicx}
\usepackage{epsfig}
\usepackage{dcolumn}
\usepackage{bm}

\newcommand{\eff}{\varepsilon}

\newcommand{\kp}{K^+}
\newcommand{\kn}{K^-}

\newcommand{\etap}{\eta^{\prime}}

\newcommand{\be}{\begin{enumerate}}
\newcommand{\ee}{\end{enumerate}}
\newcommand{\bi}{\begin{itemize}}
\newcommand{\ei}{\end{itemize}}

\newcommand{\jpsi}{J/\psi}

\newcommand{\pip}{\pi^+}
\newcommand{\pin}{\pi^-}
\newcommand{\pio}{\pi^0}
\newcommand{\g}{\gamma}
\newcommand{\ar}{\rightarrow}

\def\Journal#1&#2&#3(#4){#1{\bf #2}, #3 (#4)}

\def\bec{\begin{center}}
\def\eec{\end{center}}

\begin{document}
\title{ \bf \boldmath Measurements 
of  $\jpsi$ decays into $\phi\pio$, $\phi\eta$, and $\phi\etap$ }

\author{
M.~Ablikim$^{1}$,    J.~Z.~Bai$^{1}$,               Y.~Ban$^{11}$,
J.~G.~Bian$^{1}$,    X.~Cai$^{1}$,                  J.~F.~Chang$^{1}$,
H.~F.~Chen$^{17}$,   H.~S.~Chen$^{1}$,              H.~X.~Chen$^{1}$,
J.~C.~Chen$^{1}$
% ...
\\(BES Collaboration)\\
\vspace{0.2cm}
$^{1}${\it Institute of High Energy Physics, Beijing 100049, People's
Republic of
 China }\\
% ...
$^{11}$ {\it Peking University, Beijing 100871, People's Republic of
China}\\
% ...
$^{17}$ {\it University of Science and Technology of China, Hefei 230026,
People's Republic of
China}\\
% ...
}

\noindent\vskip 0.2cm 
\begin{abstract}
Based on $5.8 \times 10^7 \jpsi$ events detected in BESII, the
branching fractions of $\jpsi\to\phi\eta$ and $\phi\etap$ are measured
for different $\eta$ and $\etap$ decay modes. The results are
significantly higher than previous measurements.  An upper limit on
$B(\jpsi\ar\phi\pio)$ is also obtained.

\end{abstract}

\maketitle 

\section{Introduction}   \label{introd} 

The decay of the $J/\psi$ into a vector and pseudoscalar meson pair,
$J/\psi \to V P$ with $V$ and $P$ representing vector and pseudoscalar
mesons, can proceed via strong and electromagnetic reactions. A well measured set of all possible decays of
$J/\psi \to V P$ allows one to systematically study the quark gluon
contents of pseudoscalar mesons, SU(3) breaking, as well as determine
the electromagnetic and doubly suppressed OZI amplitudes in two-body
$J/\psi$ decays~\cite{theory}. MARKIII~\cite{mark2,mark3} and
DM2~\cite{dm2} measured many $J/\psi \to V P$ decays and obtained the
$\eta-\eta'$ mixing angle, the quark content of the $\eta$ and $\eta'$,
and much more.

Recently, a sample of $5.8 \times 10^7 \jpsi$ events was
accumulated with the upgraded Beijing Spectrometer
(BESII)~\cite{besii}, which offers a unique opportunity to measure precisely
the full set of $J/\psi \to V P$ decays.  In an earlier analysis
based on this data set, the branching fraction of
$\jpsi\ar\pip\pin\pio$ was measured to be
$(2.10\pm0.12)\%$~\cite{besrhopi}, which is higher than the
PDG~\cite{pdg2004} value by about 30\%.
This indicates a higher branching fraction for $\jpsi\ar\rho\pi$ than
those from older experiments~\cite{frhopi}, since the dominant dynamics
in $\jpsi \to \pip\pin\pio$ is $\jpsi\ar\rho\pi$. Therefore,
remeasuring the branching fractions of all $\jpsi\ar VP$ decay modes
becomes very important.
In this paper, $J/\psi \to \phi \pi^0$, $\phi \eta$, and $\phi \etap$ are 
studied, based on the BESII $5.8 \times 10^7 \jpsi$
events.

\section{The BES Detector}  \label{BESD} 

The upgraded Beijing Spectrometer detector (BESII) is located at the
Beijing Electron-Positron Collider (BEPC). BESII is a large
solid-angle magnetic spectrometer which is described in detail in
Ref.~\cite{besii}.  The momentum of  charged particles is determined
by a 40-layer cylindrical main drift chamber (MDC) which has a
momentum resolution of $\sigma_{p}$/p=$1.78\%\sqrt{1+p^2}$ ($p$ in
GeV/c).  Particle identification is accomplished using specific
ionization ($dE/dx$) measurements in the drift chamber and
time-of-flight (TOF) information in a barrel-like array of 48
scintillation counters. The $dE/dx$ resolution is
$\sigma_{dE/dx}\simeq8.0\%$; the TOF resolution for Bhabha events is
$\sigma_{TOF}= 180$ ps.  Radially outside of the time-of-flight
counters is a 12-radiation-length barrel shower counter (BSC)
comprised of gas proportional tubes interleaved with lead sheets. The
BSC measures the energy and direction of photons with resolutions of
$\sigma_{E}/E\simeq21\%\sqrt{E}$ ($E$ in GeV), $\sigma_{\phi}=7.9$
mrad, and $\sigma_{z}=2.3$ cm. The iron flux return of the magnet is
instrumented with three double layers of proportional counters (MUC)
that are used to identify muons.

A GEANT3 based Monte Carlo package (SIMBES) with detailed
consideration of the detector performance is used. The consistency
between data and Monte Carlo has been carefully checked in many high
purity physics channels, and the agreement is reasonable. The
detection efficiency and mass resolution for each decay mode are
obtained from a Monte Carlo simulation which takes into account the
angular distributions appropriate for the different final
states~\cite{rhopi}.

\section{analysis}
\label{analysis}
In this analysis, the $\phi$ meson is observed in its $\kp\kn$ decay
mode, and the pseudoscalar mesons are detected in the modes:
$\pio\ar\g\g$; $\eta\ar\g\g$, $\g\pip\pin$, and $\pip\pin\pio$; and
$\etap\ar\g\g$, $\g\pip\pin (\g\rho)$, and
$\pip\pin\eta~(\eta\ar\g\g)$.  Using multiple $\eta$ and $\etap$ decay
modes allows us to crosscheck our measurements, as
well as obtain higher precision. Possible final states of
$\jpsi\ar \phi\pio,~\phi\eta$, and $\phi\etap$ are then $\kp\kn\g\g$,
$\kp\kn\pip\pin\g$, and $\kp\kn\pip\pin\g\g$. Candidate events are
required to satisfy the following common selection criteria:

\begin{enumerate}
\item The events must have the correct number of
charged tracks with net charge zero. Each track must be well fitted to
a helix, originating from the interaction region of R$_{xy}<$0.02 m and
$|z| <$ 0.2 m, and have a polar angle, $\theta$, in the range
$|\cos \theta| <$ 0.8.

\item Events
should have at least the minimum number of isolated photons associated
with the different final states. 
Isolated photons are those that have
energy deposited in the BSC greater than 60 MeV, the angle
between the direction at the first hit layer of the BSC and the developing
direction of the cluster less than 30$^\circ$, and the angle between
photons and any charged tracks larger than 10$^\circ$.

\item  For each charged track in an event, $\chi^{2}_{PID}(i)$ is determined
using both $dE/dx$ and TOF information:
\begin{center}
 $\chi^{2}_{PID}(i)$=$\chi^{2}_{dE/dx}(i)$+$\chi^{2}_{TOF}(i)$
\end{center}

A charged track is identified as a $\pi$ or $K$ if its
$\chi^{2}_{PID}$ is less than those for any other assignment.  To
reject background events, two charged tracks are required to be
identified as kaons in $\jpsi\ar\phi\pio$. For the other channels, at
least one charged track must be identified as a kaon in the event
selection.

\item The selected events are fitted kinematically. The kinematic fit
  adjusts the track energy and momentum within the measured errors so
  as to satisfy energy and momentum conservation for the given event
  hypothesis. This improves resolution, selects the correct
  charged-particle assignment for the tracks, and reduces background.
When the number of photons in an event exceeds the minimum, 
all combinations  are
tried, and the combination with the smallest $\chi^{2}$ is retained.


\end{enumerate}

The branching fraction is calculated using

\begin{eqnarray*}
\lefteqn{B(\jpsi\ar \phi P) = }   \\ 
& & \frac{N_{obs}}{N_{\jpsi}\cdot
\eff\cdot B(\phi\ar \kp\kn)\cdot B(P\ar X)},
\label{forbr}
\end{eqnarray*}
where $N_{obs}$ is the number of events observed (or the upper limit),
$N_{\jpsi}$ is the number of $\jpsi$ events, $(5.77\pm
0.27)\times 10^7$, determined from the number of
inclusive 4-prong hadronic decays~\cite{fangss}, $\eff$ is the detection efficiency
obtained from Monte Carlo simulation, and $B(\phi\ar\kp\kn)$ and
$B(P\ar X)$ are the branching fractions of $\phi\ar\kp\kn$ and
pseudoscalar decays from the PDG~\cite{pdg2004}, respectively.

\subsection{\boldmath $\jpsi\ar\phi\g\g$}
Events with two oppositely charged tracks and at least two or
three isolated photons are selected.  A 4C-fit is performed to the
$K^+ K^- \gamma \gamma$ hypothesis, and $\chi^2 < 15$ is
required. To reject possible background events from
$\jpsi\ar\g\kp\kn\pio$, the 4C-fit probability for the assignment
$\jpsi\ar\kp\kn\g\g$ must be larger than that of $\kp\kn\g\g\g$.

\begin{figure}[htbp]
\centerline{\hbox{\psfig{file=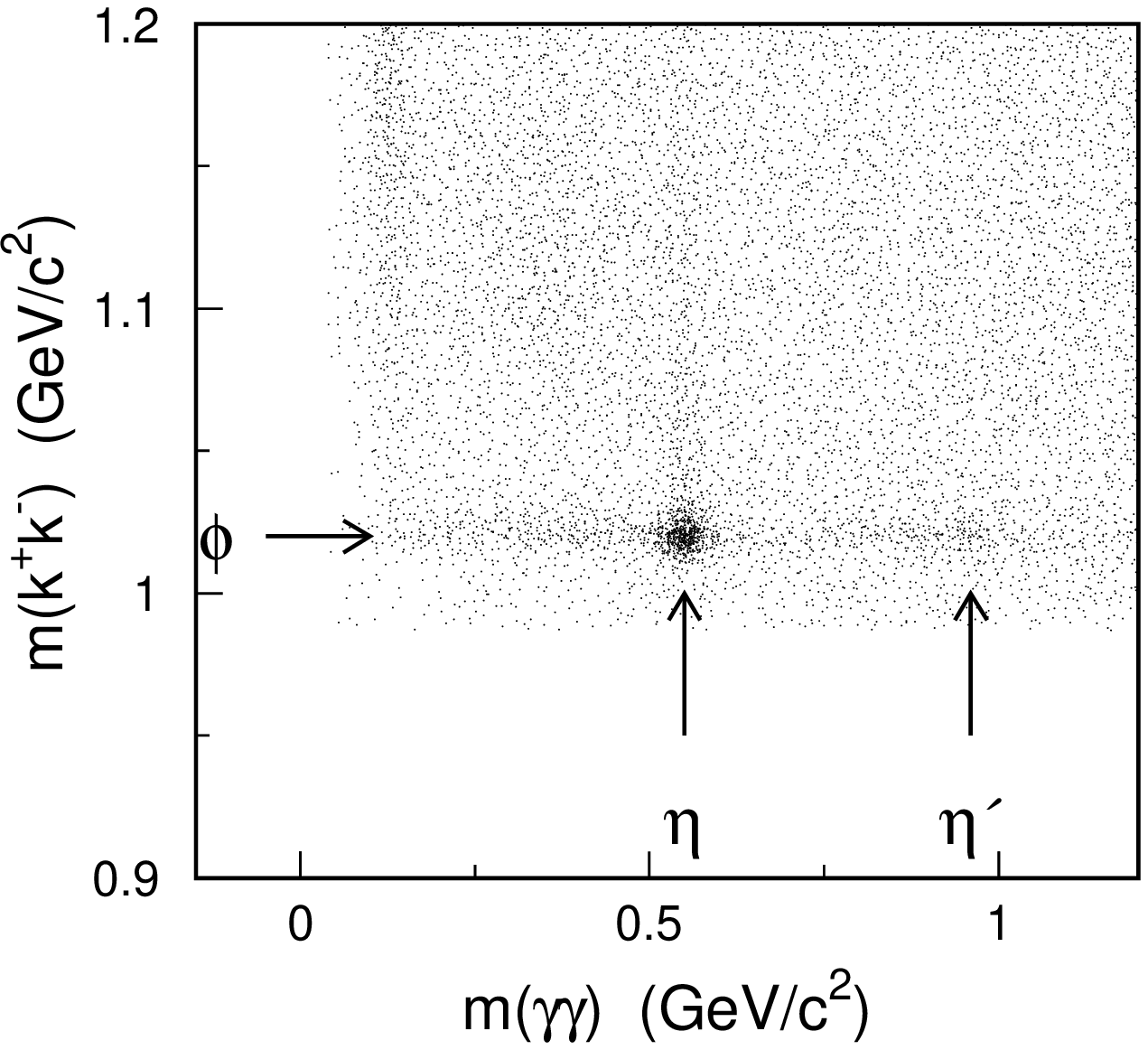,width=7cm,height=7cm}}
}
\caption{Scatter plot of $m_{\kp\kn}$ versus $m_{\g\g}$ for
  $\jpsi\ar\kp\kn\g\g$ events.}
\label{dphieta2gam}
\end{figure}

After this selection, the scatter plot (Figure~\ref{dphieta2gam})
of $m_{\kp\kn}$ versus $m_{\g\g}$ shows two clusters 
corresponding to $\phi \etap$ and $\phi\eta$, but there is no clear
accumulation of events for $\phi\pio$. To obtain the $m_{\g\g}$
distribution recoiling against $\phi$, the
$\kp\kn$ invariant mass is required to be in the $\phi$ mass region,
$|m_{\kp\kn}-1.02|<0.02$ GeV/c$^2$.

\subsubsection{$\jpsi\ar\phi\pio$}
\begin{figure}[htbp]
\centerline{\hbox{\psfig{file=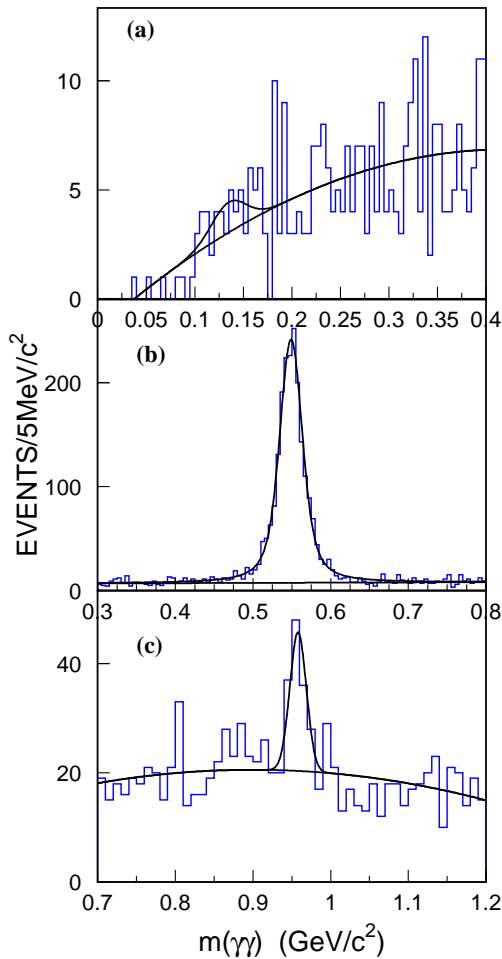,width=7.cm,height=14cm}}
}
\caption{The invariant mass distribution of $m_{\g\g}$ for
$\jpsi\ar\phi\g\g$ events. The curves are the results of the fit
described in the text. }
\label{piofit}
\end{figure}

Figure~\ref{piofit}(a) shows the $m_{\g\g}$ invariant mass distribution
after the above selection; no clear $\pio$ signal is observed. The
Bayesian method is used to determine the upper limit on the
$\jpsi\ar\phi\pio$ branching fraction. A Breit-Wigner convoluted with
a Gaussian plus a polynomial background function are used to fit the
$m_{\g\g}$ spectrum. The $\pio$ mass and width are fixed to PDG values.
The mass resolution, obtained from
Monte Carlo simulation, is 17.7 MeV/c$^2$. At the 90\% confidence
level, the number of $\phi\pio$ events is 24.  Taking into
account the detection efficiency, $(16.63\pm 0.20)\%$, the upper limit
on the branching fraction is

\begin{center}
$B(\jpsi\ar\phi\pio)<5.10\times 10^{-6}$

\end{center}

\subsubsection{$\jpsi\ar\phi\eta$}

Figure~\ref{piofit}(b) shows the $m_{\g\g}$ distribution; an $\eta$ signal is clearly seen. 
The fit of this distribution with a Breit-Wigner convoluted with
a Gaussian plus a second order 
polynomial background function 
gives $2086\pm 42 $ $\phi\eta$ events with a $\eta$ mass of $549.0\pm 0.5$ MeV/c$^2$. The background 
events, $152\pm 17$, are estimated from the $\phi$ sidebands, defined by
 0.98 GeV/c$^2 <m_{\kp\kn}<1.00$ GeV/c$^2$ and 
 1.04 GeV/c$^2<m_{\kp\kn}<1.06$ GeV/c$^2$.
After subtracting background and correcting for detection efficiency, 
$(19.98 \pm0.22)$\%,
 the $\jpsi\ar\phi\eta$ branching fraction is obtained
\begin{center}
 $B(\jpsi\ar\phi\eta)=(8.67\pm 0.19)\times 10^{-4}$,
\end{center}
where the error is statistical only.


\subsubsection{$\jpsi\ar\phi\etap$}
The distribution of $m_{\g\g}$ in $\etap$ mass region recoiling against
 the $\phi$
is shown in Figure~\ref{piofit}(c). A fit of the $\etap$ peak with a
Breit-Wigner
and a second order backgound polynomial yields $68\pm 15$ $\phi\etap$
events with the peak at $958.1\pm2.6$ MeV/c$^2$.
No obvious signal is observed  for the distribution of $m_{\g\g}$ recoiling
against $\phi$ sidebands (0.98 GeV/c$^2<m_{\kp\kn}<1.0$ GeV/c$^2$ and
1.04 GeV/c$^2<m_{\kp\kn}<1.06$ GeV/$c^2$). The detection efficiency is
$(18.57\pm 0.22)\%$,
and the corresponding branching fraction is determined to  be
\begin{center}
$B(\jpsi\ar\phi\etap)=(6.10\pm 1.34)\times 10^{-4}$,
\end{center}
where the error is only the statistical error.


\subsection{$\jpsi\ar\phi\g\pip\pin$}
\label{phietag2pi}
For $J/\psi \to \phi \eta$, $\eta \to \gamma \pi^+ \pi^-$, events with
four well-reconstructed charged tracks and at least one isolated
photon are required. To select the pions and kaons from amongst the
tracks, 4C fits are applied for one of the following three cases: (1)
if only one charged track is identified as a kaon using particle
identification, then the other charged tracks are assumed, one at a
time, to be a kaon, while the other two are assumed to be pions; (2) if
two charged tracks are identified as kaons, then the other two tracks
are assumed to be pions; (3) if three or four charged tracks are
identified as kaons, then the particle identification information is
ignored and all combinations of two kaon and two pion tracks are
kinematicaly fitted.  For each case,
the hypothesis with the smallest $\chi^{2}$ is selected.
We further require that the probability of the 4C fit for the
$\jpsi\ar\kp\kn\pip\pin\g$ assignment is larger than those of
$\kp\kn\pip\pin$ and $\kp\kn\pip\pin\g\g$.

\begin{figure}[htbp]
\centerline{\hbox{\psfig{file=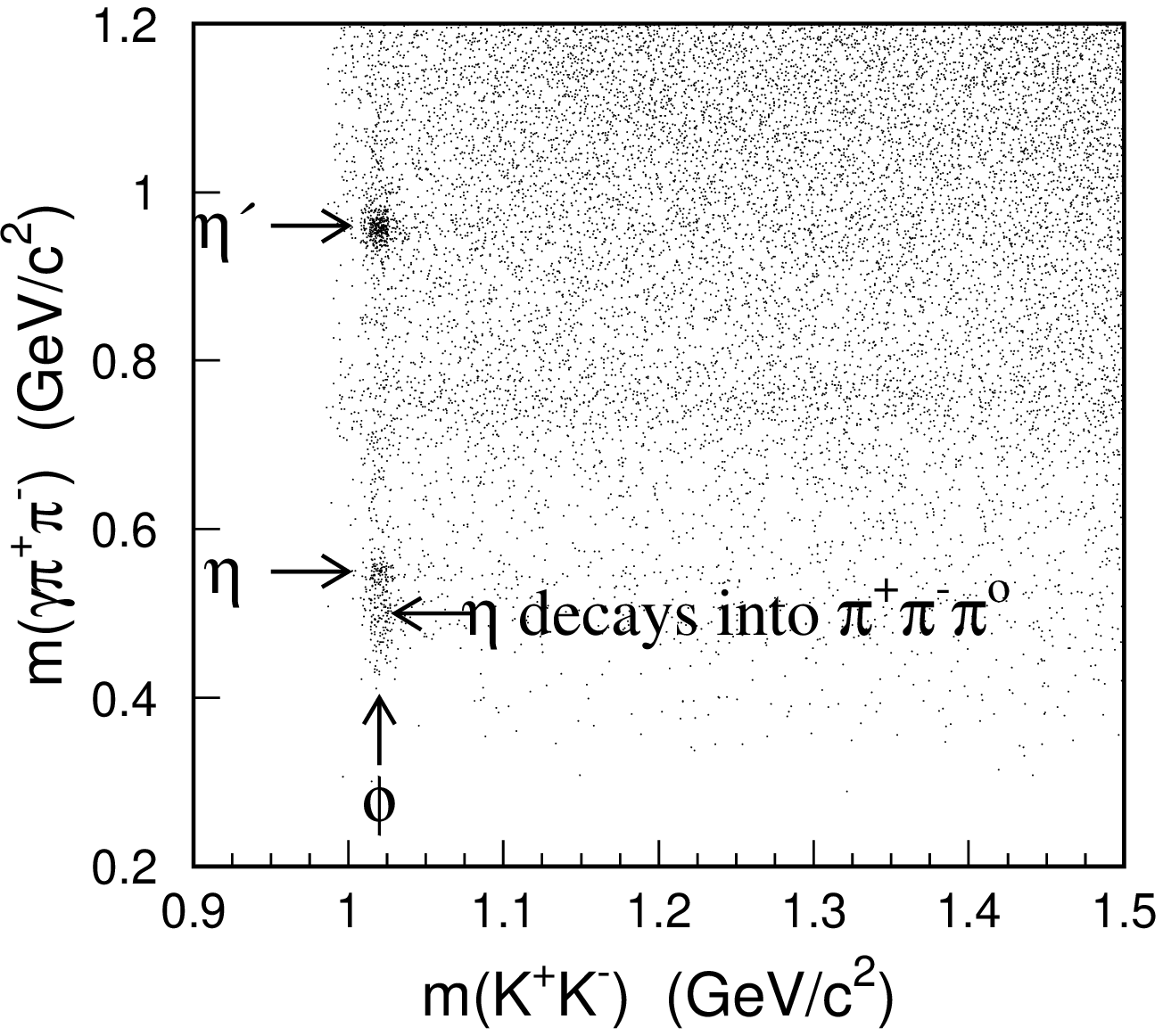,width=7cm,height=7cm}}}
\caption{Scatter plot of $m_{\kp\kn}$ versus $m_{\g\pip\pin}$ for
  $\jpsi\ar\kp\kn\pip\pin\g$ events. The
band below the $\eta$ signal
comes from $\eta\ar\pip\pin\pio$ events.}
\label{dphietag2pi}
\end{figure}

\begin{figure}[htbp]
\centerline{\hbox{\psfig{file=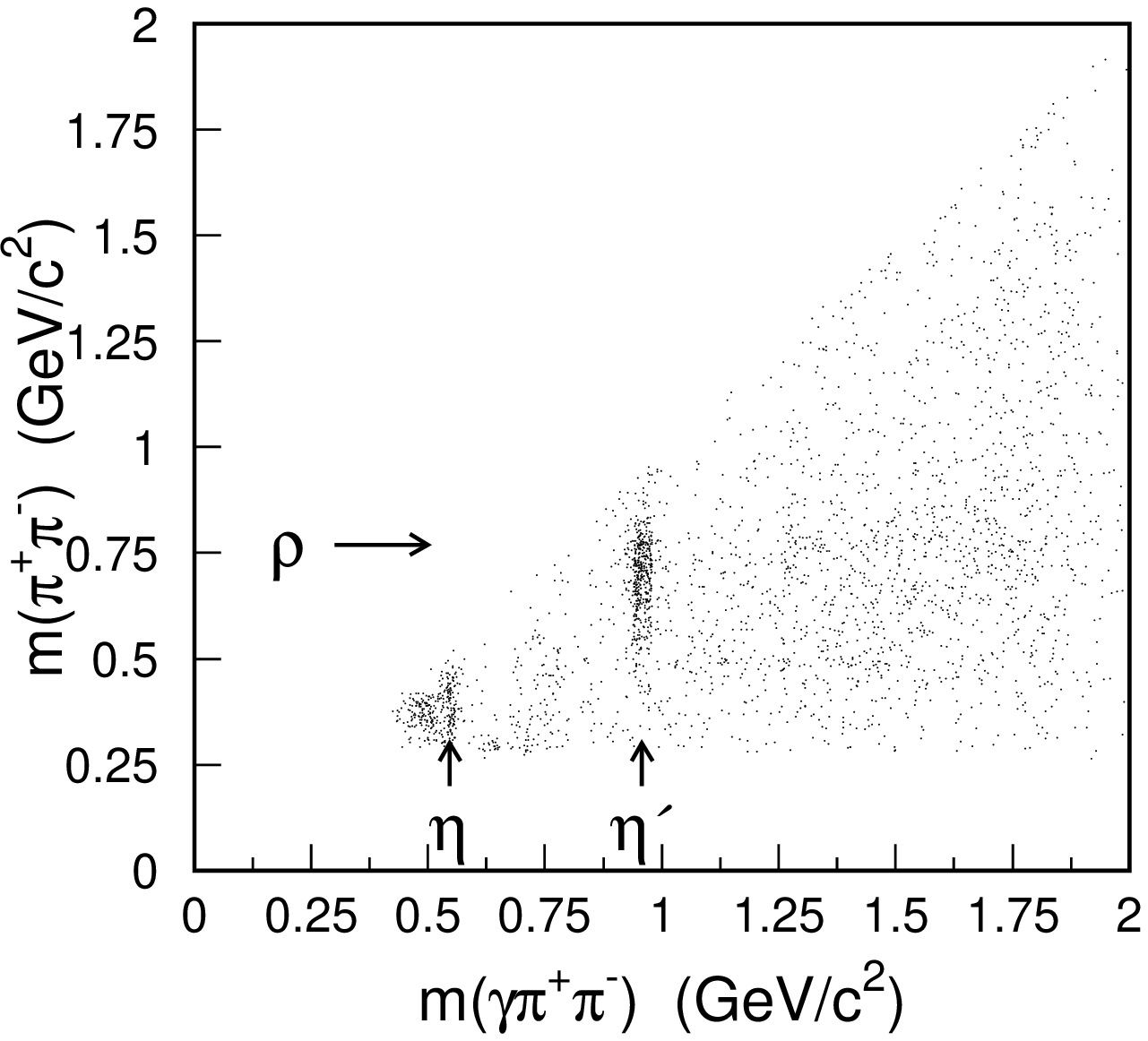,width=7cm,height=7cm}}}
\caption{Scatter plot of $m_{\pip\pin}$ versus $m_{\g\pip\pin}$ for
$\jpsi\ar\phi\g\pip\pin$
events. The $\etap$ - $\rho$ signal corresponds
to the decay $\etap\ar\g\rho$. The other cluster is from
$\eta \ar \g\pip\pin$ and $\eta\ar\pio\pip\pin$ background events. }
\label{dphietapgrho}
\end{figure}

The scatter plot of
$m_{\kp\kn}$ versus $m_{\g\pip\pin}$  is shown in
Figure~\ref{dphietag2pi}, where $\jpsi\ar\phi\eta$
and $\jpsi\ar\phi\etap$ decays are apparent. 
For the scatter plot of $m_{\pip\pin}$ versus $m_{\g\pip\pin}$, shown in
Figure~\ref{dphietapgrho},
the $\etap$ - $\rho$ signal corresponds
to the decay $\etap\ar\g\rho$. The other cluster is from
$\eta \ar \g\pip\pin$ and $\eta\ar\pio\pip\pin$ background events.

\subsubsection{$\jpsi\ar\phi\eta$}
\begin{figure}[htbp]
\centerline{\hbox{\psfig{file=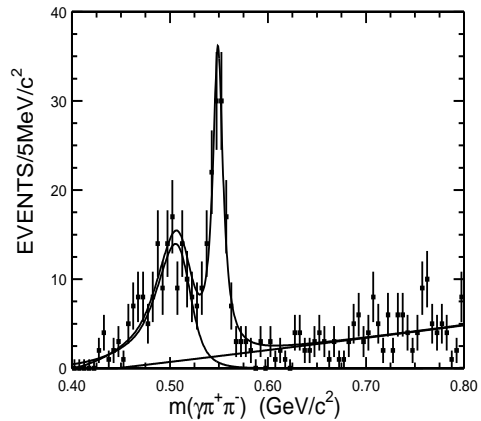,width=6.2cm,height=5.5cm}}}
\caption{Distribution of $m_{\g\pip\pin}$ for $\jpsi\ar\phi\pip\pin\g$
 events. Dots with error bars
are data, and the curves are the results of the fit described in the
text. }
\label{mgpipi}
\end{figure}

 Figure~\ref{mgpipi} shows
the $\g\pip\pin$ invariant mass recoiling against the $\phi$, defined by
$|m_{K^+K^-}-1.02|<0.02$ GeV/c$^2$. A clear $\eta$ signal is 
observed.  The peak on the left side of the $\eta$ in
Figure~\ref{mgpipi} comes from $\jpsi\ar\phi\eta~(\eta\ar\pip\pin\pio)$ 
with one photon missing; this is confirmed by Monte-Carlo simulation. 
This peak cannot be described by a simple Breit-Wigner due to its
asymmetric shape.  To obtain the shape of the peak, a Monte-carlo sample of
$\jpsi\ar\phi\eta~(\eta\ar\pip\pin\pio)$ is generated and a fit is
made to
the peak.
The $\g\pip\pin$ mass distribution is then fitted with this shape,  a Breit-Wigner
to describe
the $\eta$ signal,
and a polynomial background.  
The fit, shown in Figure~\ref{mgpipi},  
yields $134 \pm 14$  $\eta$ events with a mass at $548.9\pm0.9$ MeV/c$^2$.
The detection efficiency obtained from Monte Carlo simulation is
$(10.32\pm 0.16)\%$, and the corresponding branching fraction
is
\begin{center}
$B(\jpsi\ar\phi\eta)=(9.79\pm 1.02)\times 10^{-4},$
\end{center}
where the error is  only the statistical error.

\subsubsection{$\jpsi\ar\phi\etap$}

\begin{figure}[htbp]
\centerline{\hbox{\psfig{file=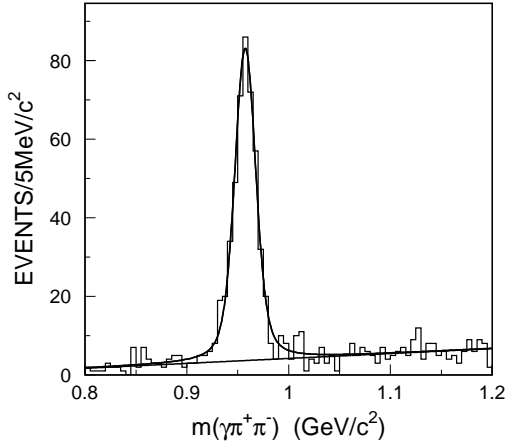,width=7cm,
height=7cm}}}
\caption{The distribution of $m_{\g\pip\pin}$ for events
of the type $\jpsi\ar\phi\rho\g$; the curves are the result of
the fit described in the text.}
\label{fitphietapg2pi}
\end{figure}

After requiring $|m_{\kp\kn}-1.02|<0.02$ GeV/c$^2$ and 0.3
GeV/c$^2<m_{\pip\pin}<0.95$ GeV/c$^2$, the distribution of
$\g\pip\pin$ invariant mass recoiling against the $\phi$ is shown in
Figure~\ref{fitphietapg2pi}; a fit with a Breit-Wigner
convoluted with a Gaussian and a second order polynomial gives $462\pm 29
$ events with a peak at $957.4\pm0.7$ MeV/c$^2$.
The detection efficiency obtained from Monte Carlo simulation
is $(9.80\pm0.16)$\%, and
the branching fraction obtained is
\begin{center}
$B(\jpsi\ar\phi\etap)=(5.64\pm0.35)\times 10^{-4}$.
\end{center}

\subsection{$\jpsi\ar\phi\pip\pin\g\g$}
For the $\eta\ar\pip\pin\pio$ case, events with four well
reconstructed charged tracks and at least two isolated photons are
selected. A 4C kinematic fit to the $\kp\kn\pip\pin\g\g$ hypothsis is
applied, as described in { Section \ref{phietag2pi} }
for $\jpsi\ar\phi\g\pip\pin$, and the case with the
smallest $\chi^2$ is selected. 

\begin{figure}[htbp]
\centerline{\hbox{\psfig{file=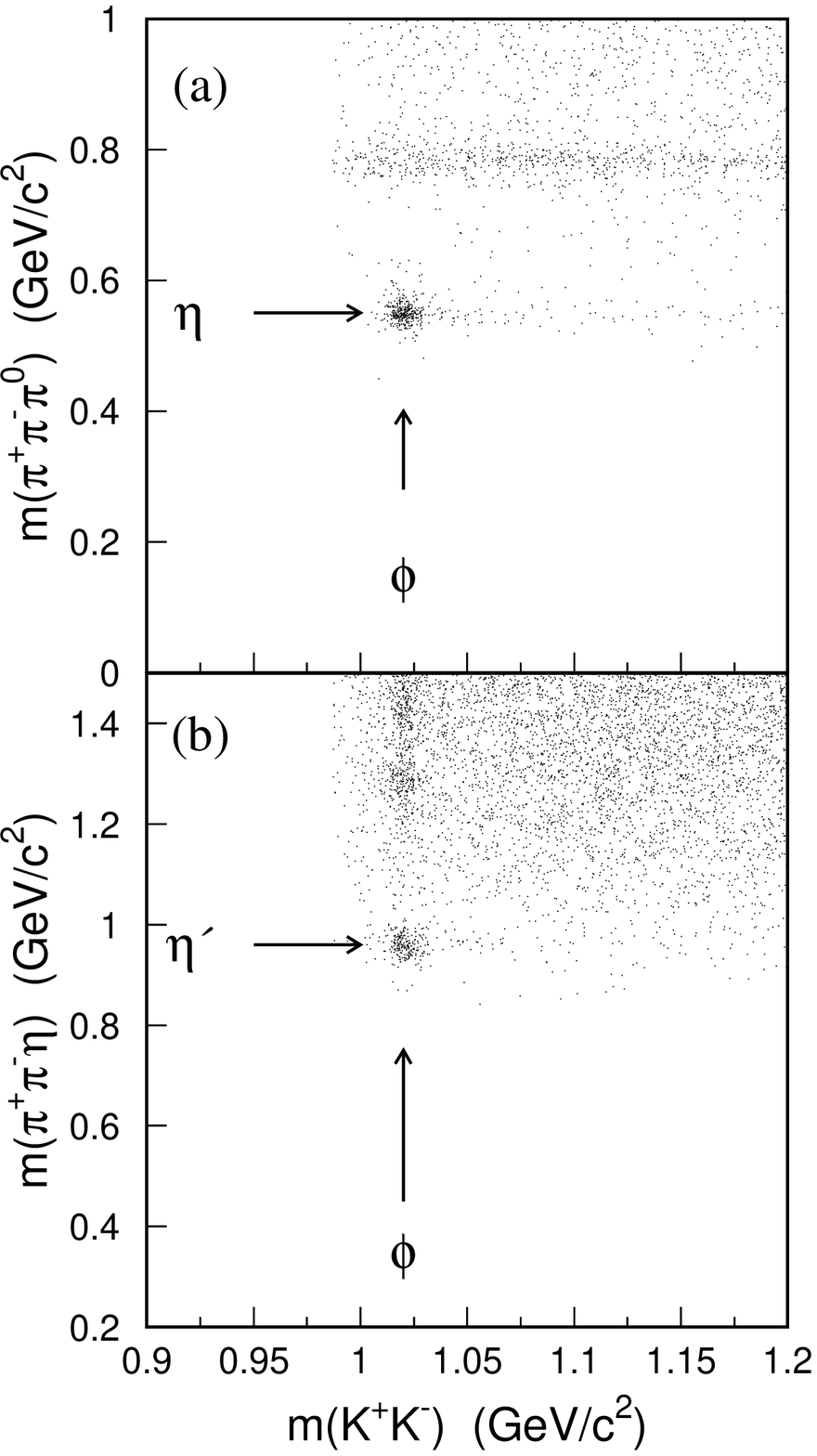,width=7cm,height=13cm}}}
\caption{Scatter plots for $m_{\kp\kn}$ versus $m_{\pip\pin\pio}$ 
and $m_{\kp\kn}$ versus $m_{\pip\pin\eta}$
for
  $\jpsi\ar\kp\kn\pip\pin\g\g$ events.} 
\label{dphieta3pi1}
\end{figure}

\begin{figure}[htbp]
\centerline{\hbox{\psfig{file=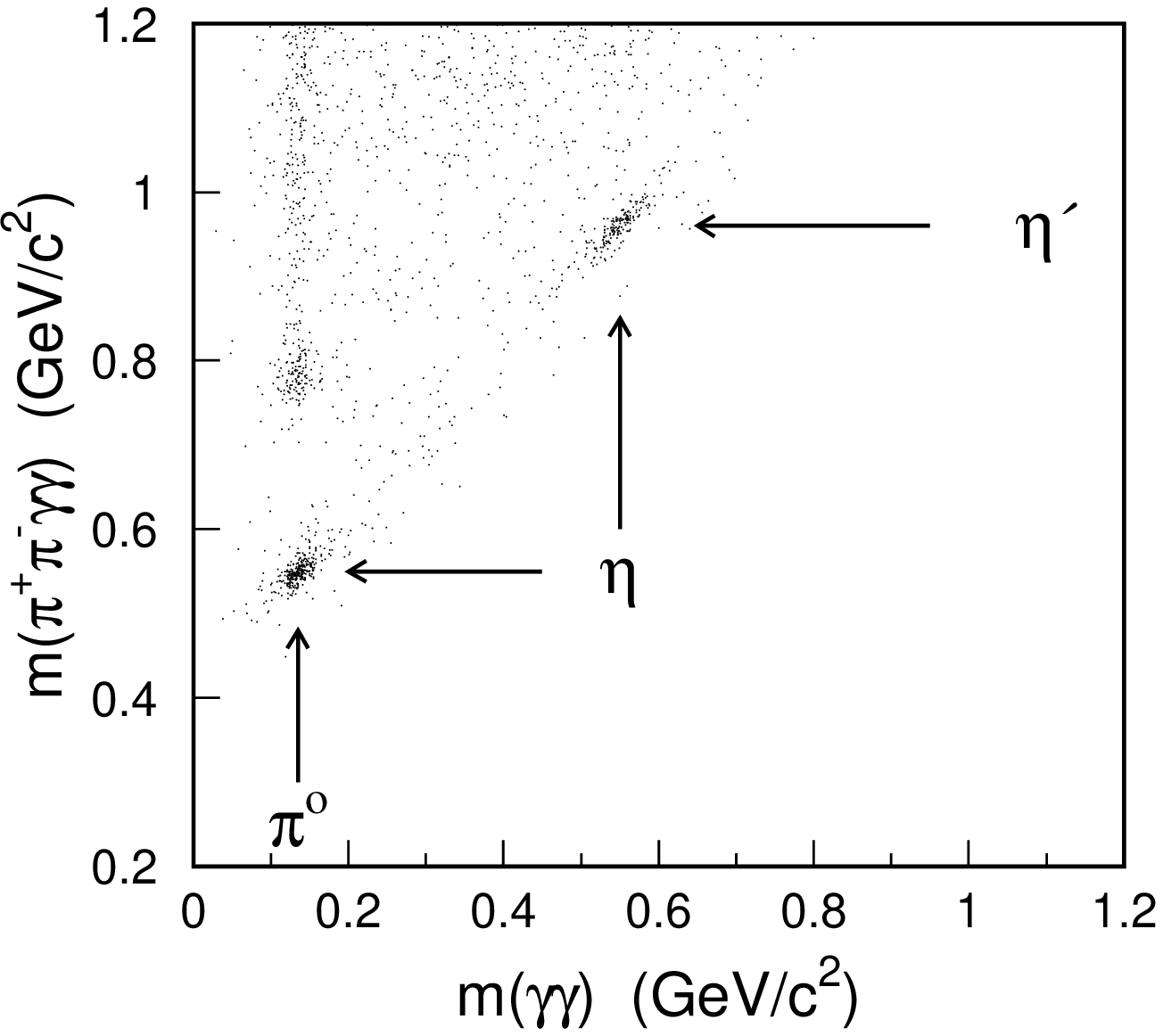,width=7cm,height=7cm}}}
\caption{Scatter plot of $m_{\g\g}$ versus $m_{\pip\pin\g\g}$ for
  $\jpsi\ar\phi\pip\pin\g\g$ candidate events.}
\label{dphietap2pieta}
\end{figure}

After the above selection and with the requirement that $m_{\g\g}$ be
consistent with a $\pio$, (0.095 GeV/c$^2$ $< m_{\gamma \gamma} <
0.175$ GeV/c$^2$), the $\jpsi\ar\phi\eta$ decay is clearly observed in
the scatter plot of $m_{\kp\kn}$ versus $m_{\pip\pin\g\g}$, shown in
Figure~\ref{dphieta3pi1}(a). Requiring 0.5 GeV/c$^2$  $< m_{\g\g} < $
0.6 GeV/c$^2$, the scatter plot in Figure~\ref{dphieta3pi1}(b) shows
clean $\phi\etap$ signals. The decays of $\eta\ar\pip\pin\pio$ and
$\etap\ar\pip\pin\eta$ are also observed in the scatter plot of
$m_{\g\g}$ versus $m_{\pip\pin\eta}$, shown in
Figure~\ref{dphietap2pieta}.


\subsubsection{$\jpsi\ar\phi\eta$}
\begin{figure}[htbp]
\centerline{\hbox{\psfig{file=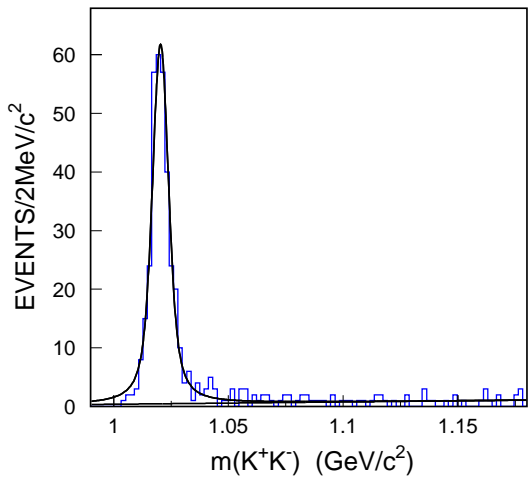,width=7.5cm,height=7.5cm}}
}
\caption{The $m_{\kp\kn}$ distribution  for
  $\jpsi\ar\kp\kn\pip\pin\pio$ events.
The curves are the results of the
fit described in the text.}
\label{m3pi}
\end{figure}

The $m_{K^+K^-}$ invariant mass spectrum recoiling against the $\eta$,
shown in Figure~\ref{m3pi}, is used to get the $\phi\eta$ signals.  A
Breit-Wigner convoluted with a Gaussian to account for the $\phi$ mass
resolution plus a second order polynomial are used to fit the
$m_{K^+K^-}$ mass distribution. A total of $350\pm$11 events with a $\phi$
mass at $1020.4\pm0.3$ MeV/c$^2$ from
$\phi$ decay are obtained in the fit, which using the detection
efficiency of $(5.81\pm0.12)$\% corresponds to a branching fraction of
\begin{center}
$B(\jpsi\ar\phi\eta)=(9.41\pm0.30)\times 10^{-4}$.
\end{center}
Here, the error is only the statistical error.

\subsubsection{\boldmath $\jpsi\ar\phi\etap$}

\begin{figure}[htbp]
\centerline{
\hbox{\psfig{file=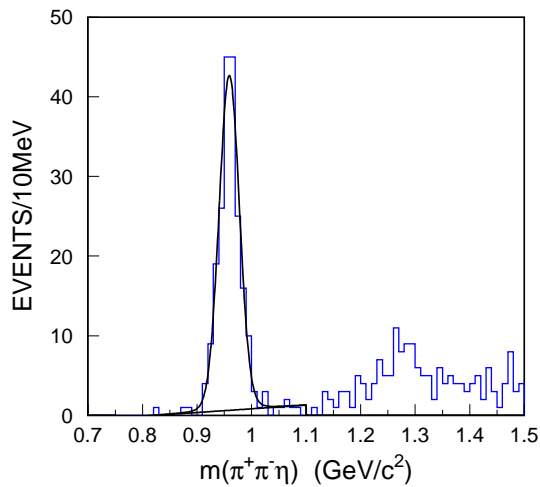,width=7.5cm,height=7.5cm}}}
\caption{The $m_{\pip\pin\g\g}$ distribution for
  $\jpsi\ar\phi\pip\pin\eta$ candidate events.
The curves are the result of
the fit described in the text.}
\label{fitphietap2pieta}
\end{figure}

After requiring $0.5 <m_{\gamma\gamma} <0.6$ GeV/c$^2$ and
$m_{\pip\pin}<0.45$ GeV/c$^2$, the 
$\pip\pin\g\g$ mass recoiling against the $\phi$($|m_{\kp\kn}-1.02|<0.02$
GeV/c$^2$), shows a clean $\etap$ peak, as seen in 
Figure~\ref{fitphietap2pieta}. No clear signal is observed for
$\phi$ sidebands (0.98 GeV/c$^2<m_{\kp\kn}<1.0$ GeV/c$^2$ and 1.04
GeV/c$^2<m_{\kp\kn}<1.06$ GeV/c$^2$). The fit of $m_{\pip\pin\g\g}$ yields
$198 \pm 12$ events with a peak at $959.2\pm1.4$ MeV/c$^2$,  and
the detection efficiency for this channel
is $(7.83\pm0.14)$\%, which gives
\begin{center}
$B(\jpsi\ar\phi\etap)=(5.11\pm0.31)\times 10^{-4}.$
\end{center}
Here, the error is  statistical only.

\subsection{Systematic Errors} \label{J-sys}
In this analysis, the systematic errors on the branching fractions mainly come 
from the following sources:
\subsubsection{MDC tracking efficiency}
 
The MDC tracking efficiency is measured in clean channels like
$\jpsi\ar\Lambda\bar{\Lambda}$ and
$\psi(2S)\ar\pip\pin\jpsi,~\jpsi\ar\mu^+\mu^-$. It is found that the Monte
Carlo simulation agrees with data within 1-2\% for each charged track. Therefore
4\%
is taken as the systematic error on the tracking efficiency for the channels
with two charged tracks and 8\% for the channels with four charged tracks
in the final states.

\subsubsection{Particle ID}

The particle identification (PID) efficiency of the kaon is studied from 
$\jpsi\ar \kp\kn\pio$ and $\jpsi\ar\phi\eta$. The results indicate that 
the kaon PID efficiency for data agrees well with that of the Monte Carlo 
simulation in the kaon momentum region less than 1.0 GeV/c.
In the analysis of $\jpsi\ar\phi\pio$, where two charged tracks are
required to be kaons, the PID efficiency difference between data and 
Monte Carlo simulation is about 3.4\%. In 
other decay modes, at least one charged track is required to be 
identified as a kaon, so the difference from PID is less than 1\%. 
Here, the difference of the PID efficiencies between data and Monte Carlo
simulation is taken as one of the systematic errors.

\subsubsection{Photon detection efficiency}

For the decay modes analyzed in this paper, one or two photons are
involved in the final states. The photon detection efficiency is 
studied from $\jpsi\ar\rho^0\pio$ in Ref.~\cite{besrhopi}. 
The results indicate
that the difference between the detection efficiency of data and MC
simulation is less than 2\% for each
photon.


\subsubsection{Kinematic fit }

The kinematic fit is a useful tool to improve resolution and reduce
background. The systematic error from the kinematic fit is studied with
the clean channel $\jpsi\ar\pip\pin\pio$, as described in
Ref.~\cite{besrhopi}. The conclusion is that the kinematic fit
efficiency difference between data and Monte Carlo simulation is about
4.1\%.  Using the same method, the decay mode
$\jpsi\ar\pip\pin\pip\pin\pio$ is also analyzed, and the kinematic fit
efficiency difference between data and Monte-Carlo is about 4.3\%. In
this paper, 5\% is conservatively taken to be the systematic error
from the kinematic fit for all analyzed decay modes.

\subsubsection{Selection criteria}

The systematic errors for additional selection criteria in specific decay
modes are estimated by comparing the efficiency difference with and
without the criterion or replacing it with a very loose
requirement. The study indicates that they are not large
compared with other systematic errors. The results are listed in
Table~\ref{toterr}

\subsubsection {Uncertainty from hadronic interaction model}

Different simulations of the hadronic interaction lead to different
efficiencies.  In this analysis, two models, FLUKA~\cite{FLUKA} and
GCALOR~\cite{GCALOR}, are used in simulating hadronic interactions
in the Monte-Carlo. The difference of the detection efficiencies from
these two Monte Carlo models is about 3\%, which is taken as the
systematic error.

\subsubsection{Uncertainty of background}

The uncertainties of the background in each channel are estimated by 
changing the background shape in the fit. The results  are listed in
Table  \ref{toterr}. 

\subsubsection{Intermediate decay branching fractions}

The branching fractions of $\phi\ar\kp\kn$ and the pseudoscalar decays
are taken from the PDG. The errors of these branching fractions are
systematic errors in our measurements and are listed in Table
\ref{toterr}.

The systematic error contributions studied above,
the error due to the uncertainty of the number of $\jpsi$ events, and
the statistical error of the Monte-Carlo samples are all listed in 
Table \ref{toterr}. The total systematic error is the sum of them
added  in 
quadrature.

\begin{table*}[htpb]
\caption{Summary of systematic errors. }
\begin{center}
\begin{tabular}{ l|c|c|c|c|c|c|c}
\hline
\hline
$\jpsi\ar$ &$\phi\pio$ &\multicolumn{3}{|c|}{$\phi\eta$} &
\multicolumn{3}{|c}{$\phi\etap$} \\
\hline
Final states &$\kp\kn\g\g$ &$\kp\kn\g\g$ &$\kp\kn\pip\pin\g$
&$\kp\kn\pip\pin\g\g$&
$\kp\kn\g\g$ &$\kp\kn\pip\pin\g$ &$\kp\kn\pip\pin\g\g$\\

\hline
Error Sources & \multicolumn{7}{|c}{Relative Error~(\%)}\\
\hline

MDC tracking & 4  &  4  &  8   &  8  &  4  &  8 &  8 \\
Particle ID &3.4 & $<$1& $<$1  & $<$1 & $<$1 & $<$1 & $<$1\\

Kinematic fit& 5  &  5  &  5   &  5  &  5  &  5 &  5 \\

Photon efficiency &  4  &  4   &  2  &  4  &  4 & 2 & 4 \\
Selection criteria & 2.4 & 2.4 & 2.8 &1& 2.4&2.9 &2.2\\

MC sample &1.2 &1.2 & 1.5  & 2.1 & 1.2 & 1.6 & 1.8\\

Hadronic interaction model & 3 & 3 & 3  & 3 & 3 & 3 & 3\\

Background uncertainty &16.7 &3.9 & 1.5  & 3.4 &1.5 &2.0 &1.5 \\

Intermediate decays & 1.2 &1.4 & 2.7  & 2.2 & 6.7 & 3.6 & 3.6\\

Total $\jpsi$ events& 4.7 & 4.7 & 4.7  & 4.7 & 4.7 & 4.7 & 4.7\\
\hline
Total systematic error & 19.7 & 10.7 & 12.0  & 12.6 & 12.0 & 12.4
&12.7\\
\hline
\hline
\end{tabular}
\label{toterr}
\end{center}
\end{table*}

\section{Results and Discussion}
The branching fractions of $\jpsi$ decaying into $\phi\pio$,
$\phi\eta$, and $\phi\etap$, measured into different final states,
are listed in Table \ref{brpv}.  The average value is the weighted
mean of the results from the different decay modes, and the PDG value
is the world average taken from Ref.~\cite{pdg2004}.  The world
averages mainly come from MarkIII and DM2. The results obtained here
are not in good agreement with previous measurements. Just as for
the branching fraction of $\jpsi\ar\pip\pin\pio$, the branching
fraction of $\jpsi\ar\phi\eta$ and $\phi\etap$ are higher than those in
the PDG.

In this paper, we measured the branching fractions of $\jpsi$ decays
into $\phi$ plus a pseudoscalar. The three branching fractions are not
sufficient for a detailed study of pseudoscalar mixing, SU(3)
breaking, and the contribution from doubly suppressed OZI processes
using the phenomenological model in Ref.~\cite{theory}. However the
inconsistency between the results from BESII and those from former
measurements emphasize the importance for such a study. After
measuring the other decay modes of $\jpsi\ar VP$, such as
$\jpsi\ar\omega\pio$, $\omega\eta$, $\omega\etap$, $\rho\eta$, $\rho\etap$,
and $K^*K$, it will be important to extract physics with all the relevant
measurements again.

\begin{table}[htpb]
\caption{Branching fractions of $\jpsi\ar\phi\pio$, $\phi\eta$, and $\phi\etap$.}
\begin{center}
\begin{tabular}{ l |c|c }
\hline
\hline
$\jpsi\ar$          & Final states &Branching Fraction
($\times 10^{-4}$) \\
\hline
$\phi\pio$& $\kp\kn\g\g$ & $<0.064$ (C.L. 90\%)~\cite{foot}
 \\

\hline
             &$\kp\kn\g\g$ & 8.67$\pm$0.19$\pm$0.93 \\

              & $\kp\kn\pip\pin\g$ & 9.79$\pm$1.02$\pm$1.17\\
$\phi\eta$    & $\kp\kn\pip\pin\g\g$ & 9.41$\pm$0.30$\pm$1.19 \\
              & Average   &8.99$\pm$0.18$\pm$0.89 \\
              & PDG       & $6.5\pm0.7$ \\
\hline
              &$\kp\kn\g\g$ & 6.10$\pm$1.34$\pm$0.73\\

$\phi\etap$  & $\kp\kn\pip\pin\g$ & 5.64$\pm$0.35$\pm$0.70 \\

              & $\kp\kn\pip\pin\g\g$ & 5.11$\pm$0.31$\pm$0.65\\
              & Average              & 5.40$\pm$0.25$\pm$0.56\\
              & PDG                  & $3.3\pm 0.4$ \\
\hline
\hline
\end{tabular}
\end{center}
\label{brpv}
\end{table}

\acknowledgments

   The BES collaboration thanks the staff of BEPC and computing center for their hard efforts.
This work is supported in part by the National Natural Science Foundation
of China under contracts Nos. 19991480, 10225524, 10225525, the Chinese
Academy
of Sciences under contract No. KJ 95T-03, the 100 Talents Program of CAS
under Contract Nos. U-11, U-24, U-25, and the Knowledge Innovation Project
of
CAS under Contract Nos. U-602, U-34 (IHEP); by the National Natural Science
Foundation of China under Contract No. 10175060 (USTC), and
No. 10225522 (Tsinghua University) and by the Department
of Energy under Contract No. DE-FG03-94ER40833 (U Hawaii).


\begin{thebibliography}{120}
\bibitem{theory} H. E. Haber, J. Perrier, Phys. Rev. D~{\bf 32}, 2961 (1985). 
\bibitem{mark2} R. M. Baltrusaitis {\it et al.}, Phys. Rev. D~{\bf 32}, 2883 (1985).
\bibitem{mark3} D. Coffman {\it et al.}, Phys. Rev. D~{\bf 38}, 22695 (1988).
\bibitem{dm2}J. Jousset {\it et al.}, Phys. Rev. D~{\bf 41}, 1389 (1990).
\bibitem{besii} J. Z. Bai {\it et al.}, Nucl. Instrum. Methods
A~{\bf 458}, 627 (2001).

\bibitem{besrhopi} J. Z. Bai {\it et al.}, Phys. Rev. D~{\bf 70}, 012005 (2004).
\bibitem{pdg2004} S. Eidelman {\it et al.}~(Particle Data
Group), Phys. Lett. B~{\bf 592}, 1 (2004), and references therein.


\bibitem{frhopi}
J. J. Aubert {\it et al.}, Phys. Rev. Lett. {\bf 33}, 1404 (1974);
J. E. Augustin {\it et al.}, Phys. Rev. Lett. {\bf 33}, 1406 (1974);
B. Jean-Marie {\it et al.}, Phys. Rev. Lett. {\bf 36}, 291 (1976);
 W. Braunschweig {\it et al.}, Phys. Lett. {\bf 63B}, 487 (1976);
W. Bartel et al., Phys. Lett. ~{\bf 64B}, 483 (1976);
PLUTO collaboration, Phys. Lett. {\bf 72B}, 493 (1978);
 DASP collaboration, Phys. Lett. {\bf 74B}, 292 (1978);
D. Coffman et al., Phys. Rev. D~{\bf 38}, 2695 (1988);
 J. Z. Bai et al., Phys. Rev. D~{\bf 54}, 1221 (1996).


\bibitem{rhopi} The angular distribution is described by
\begin{center}
${\frac{d^3\sigma}{d\cos\theta_V d\cos\theta_1 d\phi_1}}
={\sin^2\theta_1[1+\cos^2\theta_V+\sin^2\theta_V\cos(2\phi_1)]}$
\end{center}
where $\theta_V$ is the angle between the vector meson and the positron
direction. $\theta_1$ and $\phi_1$ describe the decay products of the vector
meson in its helicity frame.
For $\phi\ar\kp\kn$, $\theta_1$ and $\phi_1$ are the polar
and azimuthal angles of the
momentum of  $K$ with respect to the helicity direction of the $\phi$.




\bibitem{fangss} S. S. Fang {\it et al.},
High Energy Phys. Nucl. Phys. {\bf 27}, 277 (2003) (in Chinese).




\bibitem{FLUKA} K. Hanssgen, H.-J.Mohring and J. Ranft, Nucl. Sci. Eng.
{\bf 88}, 551 (1984);
 J. Ranft and S. Ritter, Z. Phys. C~{\bf 20}, 347 (1983);
 A. Fasso et al., FLUKA 92, Proceedings of the Workshop on Simulating
Accelerator Radiation Environments, Santa Fe, 1993.
\bibitem{GCALOR}C. Zeitnitz and
T. A. Gabriel, Nucl. Instrum. Methods A~{\bf 349}, 106 (1994).


\bibitem{foot} {To conservatively estimate the upper limit, the result
obtained from
formula in Section.~\ref{analysis} is corrected by dividing a factor $(1- \delta_{sys})$. Here,
$\delta_{sys}$ is the systematic error for this decay mode.}



\end{thebibliography}
\end{document}